\documentclass[conference,twoside,letterpaper]{IEEEtran}
\IEEEoverridecommandlockouts
\ifCLASSINFOpdf
  \usepackage[pdftex]{graphicx}
\else
\fi

\usepackage[cmex10]{amsmath}
\usepackage{cite}
\usepackage{mathtools}
\usepackage{flushend}
\usepackage{mathdots}
\usepackage{color}
\begin{document}
\title{Modeling and Efficient Cancellation of Nonlinear Self-Interference in MIMO Full-Duplex Transceivers}

\author{\IEEEauthorblockN{Lauri Anttila\IEEEauthorrefmark{1},
Dani Korpi\IEEEauthorrefmark{1},
Emilio Antonio-Rodr\'{i}guez\IEEEauthorrefmark{2},
Risto Wichman\IEEEauthorrefmark{2}, and
Mikko Valkama\IEEEauthorrefmark{1}}
\\
\IEEEauthorblockA{\IEEEauthorrefmark{1}Department of Electronics and Communications Engineering, Tampere University of Technology, Finland}
\IEEEauthorblockA{\IEEEauthorrefmark{2}Aalto University School of Electrical Engineering, Finland\\ e-mail: \{lauri.anttila,dani.korpi,mikko.e.valkama\}@tut.fi, \{emilio.antoniorodriguez,risto.wichman\}@aalto.fi} %\vspace{-5mm}
%e-mail: \{lauri.anttila,dani.korpi,mikko.e.valkama\}@tut.fi, \{emilio.antoniorodriguez,risto.wichman\}@aalto.fi
\thanks{The research work leading to these results was funded by the Academy of Finland (under the projects \#259915, \#258364 "In-band Full-Duplex MIMO Transmission: A Breakthrough to High-Speed Low-Latency Mobile Networks"), the Finnish Funding Agency for Technology and Innovation (Tekes, under the project "Full-Duplex Cognitive Radio"), the Internet of Things program of DIGILE (Finnish Strategic Centre for Science, Technology and Innovation in the field of ICT; funded by Tekes), the Linz Center of Mechatronics (LCM) in the framework of the Austrian COMET-K2 programme, and Emil Aaltonen Foundation.} 
}
\maketitle

\begin{abstract}
This paper addresses the modeling and digital cancellation of self-interference in in-band full-duplex (FD) transceivers with multiple transmit and receive antennas. The self-interference modeling and the proposed nonlinear spatio-temporal digital canceller structure takes into account, by design, the effects of I/Q modulator imbalances and power amplifier (PA) nonlinearities with memory, in addition to the multipath self-interference propagation channels and the analog RF cancellation stage. The proposed solution is the first cancellation technique in the literature which can handle such a self-interference scenario. It is shown by comprehensive simulations with realistic RF component parameters and with two different PA models to clearly outperform the current state-of-the-art digital self-interference cancellers, and to clearly extend the usable transmit power range.
\end{abstract}

\IEEEpeerreviewmaketitle

\section{Introduction}

Full-duplex communications using the same carrier for simultaneous transmission and reception is a novel paradigm in the research on wireless communications \cite{Choi10,Jain11,Duarte12,Bharadia13,Everett13}. It would enable the full utilization of the available resources for both the transmitted and received waveforms, thus potentially doubling the achievable spectral efficiency. This is obviously a very appealing prospect, since the spectral resources are already very limited, and thus more and more users must be served without significantly increasing the amount of utilized resources. In-band full-duplex radio technology is also receiving increasing interest in currently emerging 5G mobile cellular radio technology research, see, e.g., \cite{Sabharwal14,Chih14,Hong14}.

However, the obvious problem with simultaneous transmission and reception at the same carrier is the own transmit signal, which is coupled back to the receiver and thereby acts as a powerful source of interference. This so-called \textit{self-interference} (SI) poses a fundamental challenge for in-band full-duplex transceivers, and its suppression is necessary in order to make this concept feasible in the first place \cite{Choi10,Jain11,Duarte12}. In principle, the SI signal can be cancelled by merely subtracting the transmitted signal from the received signal, as it is obviously known within the device \cite{Choi10,Jain11,Duarte12}. In practice, however, this requires knowledge about the coupling channel between the transmitter and receiver, as well as additional circuitry in the radio frequency (RF) front-end. Thus, the increased spectral efficiency comes at a cost because the effective SI coupling channel must be estimated and additional components and processing must be then used to cancel the SI signal.

One significant obstacle for efficient SI cancellation is caused by the analog circuit nonidealities, which distort the transmit signal \cite{Korpi13,Anttila13,Ahmed13,Bharadia13,Korpi132}. This means that it is not possible to accurately regenerate the received SI signal with only linear processing methods. This, on the other hand, results in less attenuation for the SI, and in some cases decreases the performance of the full-duplex transceivers below acceptable levels. One such analog nonideality is the nonlinearity of the components, which has been widely analyzed in recent literature \cite{Korpi13,Anttila13,Ahmed13,Bharadia13,Korpi132}. It has been shown that nonlinear distortion of the SI signal can result in a large performance loss if only linear SI cancellation techniques are used \cite{Korpi13}. The nonlinearity of the transmitter power amplifier (PA) has been observed to be especially harmful \cite{Anttila13}. This problem cannot be avoided in practical devices, since the PA needs to be driven close to saturation for power-efficient operation \cite{Raab_overview}. To combat this issue in full-duplex devices, several nonlinear SI cancellation techniques have already been proposed \cite{Anttila13,Ahmed13,Bharadia13}. With these methods, it is possible to accurately regenerate also a nonlinearly distorted SI signal, and thus cancel it more efficiently.

Another significant analog impairment in most wireless radio transceivers is I/Q imaging \cite{Anttila11}, which is caused by the mismatches between the I- and Q-branches of the transmitter and receiver chains. This is a widespread issue in wireless communications since nowadays most transceiver structures utilize I/Q processing. It has been shown that I/Q imaging is especially harmful in in-band full-duplex transceivers because the power of the SI signal can be significantly higher than that of the received signal of interest \cite{Korpi14}. This means that also the power of the I/Q image component can be very high in comparison to the signal of interest, possibly resulting in a low signal-to-interference-plus-noise ratio. However, in \cite{Korpi14} it was shown that it is possible to attenuate also this SI image component in the digital domain very efficiently by utilizing widely-linear signal models and processing.

In this paper, we consider both of these sources of nonideality, and develop a complete digital cancellation solution that is able to regenerate a nonlinearly distorted SI signal even when there is I/Q imbalance in the transmitter chain. In addition, for generality, the canceller is derived for a MIMO full-duplex transceiver with multiple simultaneously operating transmit and receive antennas. \emph{To the best of our knowledge this is the first time this type of a digital cancellation solution has been developed and reported in the general MIMO full-duplex transceiver scenario, and is also the first reported digital cancellation solution being able to handle joint effects of I/Q imaging and PA nonlinearities.} Since I/Q imaging and nonlinear distortion are typically the dominant sources of distortion in the analog domain \cite{Korpi14,Korpi13}, the proposed solution can be expected to provide a significant improvement in the final SINR, while having increased robustness against heavily nonideal RF components. This is verified and demonstrated in this article with comprehensive waveform simulations. We also wish to emphasize that the nonlinearities of the transmitter, generally dictated by the transmitter emission mask and limits, are particularly challenging in the full-duplex radio context, stemming from the high sensitivity requirements of the simultaneously operating receiver chain(s). Hence, even if the transmitter nonlinearities are at levels well conforming to the typical emission requirements, the remaining nonlinearities are still substantial from the perspective of the sensitive receiver. This is the key motivation behind this work.

The rest of this paper is organized as follows. In Section~\ref{sec:modeling}, the baseband signal model for the nonlinear self-interference is derived. The proposed canceller structure and parameter estimation are introduced in Section~\ref{sec:cancellation}, where we also discuss ways to decrease the complexity of the canceller. Section~\ref{sec:simulations} presents comprehensive simulation results on the performance of the proposed solution in comparison with state-of-the-art reference techniques. Finally, Section~\ref{sec:conclusions} concludes the article.

\section{Baseband Equivalent Signal Modeling}
\label{sec:modeling}

In this section we build a complete self-interference channel model for a multiple antenna full-duplex device, including the effects of transmitter impairments (PA nonlinearity and I/Q modulator imbalance), the linear MIMO self-interference channel, and analog RF SI cancellation. In this study, the receiver I/Q imbalances are assumed to be calibrated, using for example the blind compensation techniques from \cite{Anttila11}. A residual receiver I/Q imbalance level is anyway included in the simulations. 

\begin{figure*}[!t]
\centering
\includegraphics[width=0.9\textwidth]{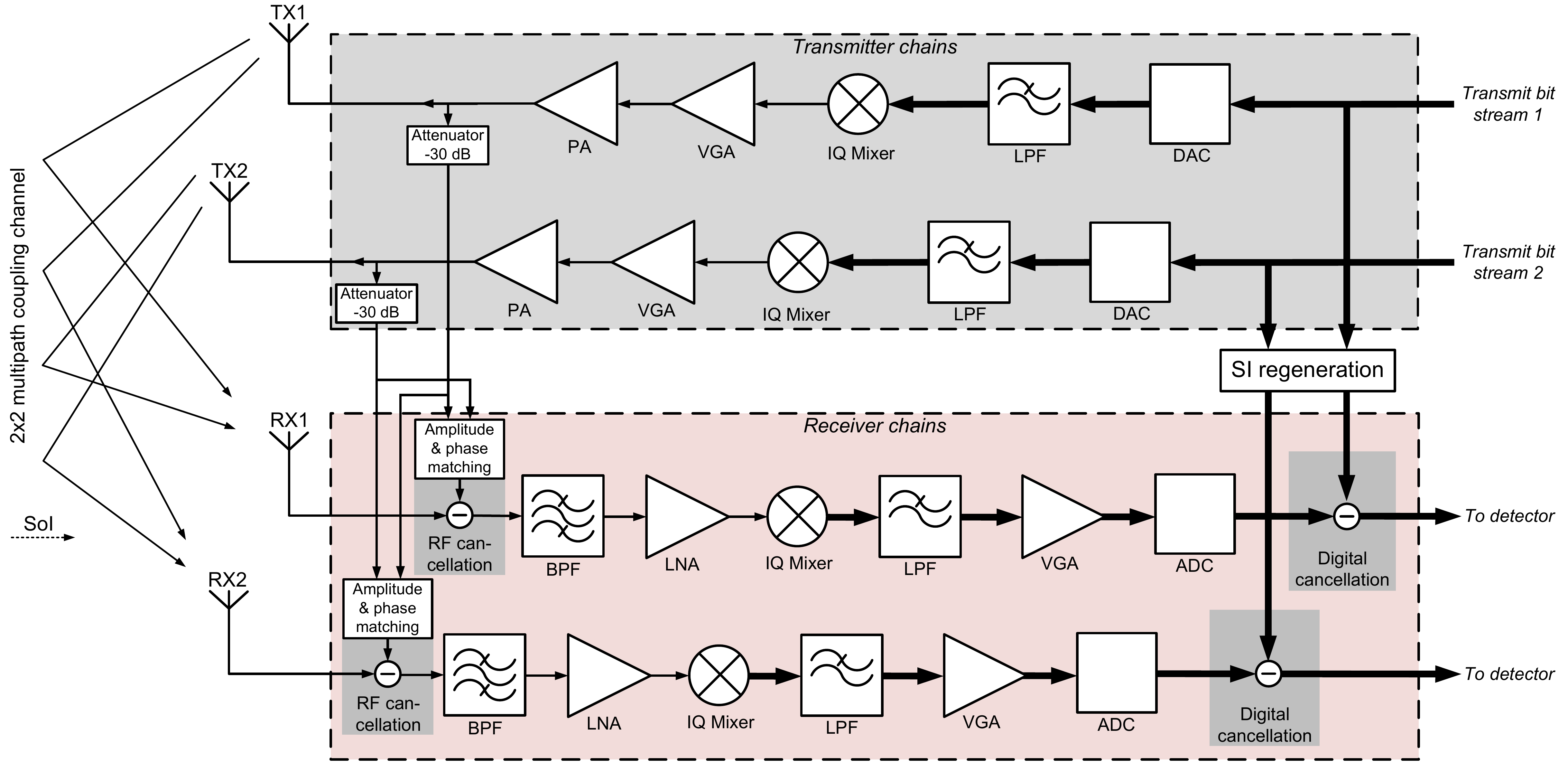}
\caption{Block diagram of the considered $2\times2$ full-duplex transceiver.}
\label{fig:block_diagram}
\end{figure*}

The proposed model is linear-in-parameters, thus lending itself well to simple parameter estimation schemes. The dimensionality of the model is increased compared to a serial, decoupled model, but with the benefit of simpler estimation. Some approximate models are also presented, based on practical insight of the transmitter impairments and their relative strengths, in order to reduce the complexity. For notational simplicity, the actual received signal of interest (SoI) and additive noise are not included in the following presentation but are naturally present in the performance simulations. An illustration of the considered full-duplex MIMO transceiver is given in Fig. \ref{fig:block_diagram} with two TX and RX antennas. All the modeling, estimation, and cancellation algorithm developments are, however, carried out for a general $N_T \times N_R$ transceiver.

\subsection{Power Amplifier and I/Q Modulator Models}
The baseband signal of transmitter $j$ ($j=1,2,\ldots ,N_T$) is denoted by $x_j(n)$. The output signal of the I/Q modulator model (frequency-independent model for simplicity) is \cite{Anttila11}
\begin{align}
x_{j}^{IQM}(n)=K_{1,j}x_j(n)+K_{2,j}x_{j}^{*}(n)
\label{eq:iq_model}
\end{align}
with $K_{1,j}={\scriptstyle 1\!/\!_2}(1+g_j\exp(j{\varphi }_{j}))$ and $K_{2,j}={\scriptstyle 1\!/\!_2}(1-g_j\exp(j{\varphi }_{j}))$, where $g_j,{\varphi }_{j}$ are the gain and phase imbalance parameters of transmitter $j$. Notice that for any practical transmitter front-end $| K_{1,j} |\gg | K_{2,j} |$. The strength of the induced I/Q image component, represented by the conjugated signal term in \eqref{eq:iq_model}, is typically characterized with the image rejection ratio (IRR) as $10\log_{10}({|K_{1,j}|^2}/{|K_{2,j}|^2})$.

The assumed power amplifier model is parallel Hammerstein (PH) with polynomial branch nonlinearities and FIR branch filters, given for transmitter $j$ as
\begin{align}
x_j^{PA}(n) = \sum_{\substack{
   p = 1 \\
   p \text{ } \mathit{odd}
  }}^P
 \sum_{m = 0}^M h_{p,j}(m) \psi_p(x_j^{IQM}(n-m)) \text{.}
 % {{|x_{j}^{IQM}(n-m)|}^{p-1}}x_{j}^{IQM}(n-m) 
\label{eq:pa_model}
\end{align}
Here, the basis functions are defined as 
\begin{align}
	\psi_p(x(n)) = |x(n)|^{p-1} x(n) = x(n)^{\frac{p+1}{2}} x^*(n)^{\frac{p-1}{2}}
\label{eq:ph_basis_functions}
\end{align}
and $h_{p,j}(n)$ denote the FIR filter impulse responses of the PH branches for transmitter $j$, while $M$ and $P$ denote the memory depth and nonlinearity order of the PH model, respectively \cite{Ding04,Isaksson06,Anttila10}. The PH nonlinearity is a widely used nonlinear model for direct as well as inverse modeling of power amplifiers and has been observed, through RF measurements, to characterize the operation of various power amplifiers in an accurate manner \cite{Ding04,Isaksson06,Anttila10,Ku03} .

Inserting now (\ref{eq:iq_model}) and (\ref{eq:ph_basis_functions}) into (\ref{eq:pa_model}) we obtain, after some straightforward but tedious algebra,
\begin{align}
x_j^{PA}(n) = \sum_{\substack{
   p = 1 \\
   p \text{ } \mathit{odd}
  }}^P
 \sum_{q=0}^p \sum_{m=0}^M h_{p,j}^{(q,p-q)}(m)\times \nonumber\\
 {x_j}{(n-m)}^q x_j^*{(n-m)}^{p-q}
\label{eq:pa_iq_cascade}
\end{align}
where $h_{p,j}^{(q,p-q)}$ are the coefficients for the basis function of the form $x^q x^{*p-q}$. These are, for order $p=1$ given as
\begin{align}
h_{1,j}^{(1,0)}(m) &= K_{1,j}h_{1,j}(m) \nonumber\\
h_{1,j}^{(0,1)}(m) &= K_{2,j}h_{1,j}(m) \nonumber
\end{align}
and for $p=3$ as
\begin{align}
h_{3,j}^{(3,0)}(m) &= K_{1,j}^{2}K_{2,j}^{*}{h_{3,j}}(m) \nonumber\\
h_{3,j}^{(2,1)}(m) &= ({|{K_{1,j}}|}^2 K_{1,j}+2{|K_{2,j}|}^2 K_{1,j})h_{3,j}(m) \nonumber\\
h_{3,j}^{(1,2)}(m) &= ({2|{K_{1,j}}|}^2 K_{2,j}+{|K_{2,j}|}^2 K_{2,j})h_{3,j}(m) \nonumber\\
h_{3,j}^{(0,3)}(m) &= K_{1,j}^{*}K_{2,j}^{2}{h_{3,j}}(m) \nonumber
\end{align}
Higher orders follow similarly but are not written out explicitly due to space constraints.

In general, with the above cascaded modeling approach for I/Q modulator and PA impairments, there are $p+1$ distinct filters for $p$th order nonlinearity, i.e., one for each distinct basis function of order $p$. However, many of the terms arising from cascading the impairments (such as $K_{1,j}^{*}K_{2,j}^2{h_{3,j}}(m)$) are so small that they can be neglected with very little effect on overall modeling accuracy. This will reduce the computational cost of such modeling and the corresponding cancellation. This is explored further in Section \ref{sec:simplified}.

\subsection{MIMO Channel and RF Canceller Model}
We denote the actual MIMO propagation channel impulse response from TX antenna $j$ to RX antenna $i$ by $c_{ij}(l),~l=0,1,\ldots ,L,$, with $L$ denoting the effective delay spread of the self-interference channel. The received SI signal at RX antenna $i$ ($i=1,2,\ldots ,N_R$) can now be written as
\begin{align}
z_i(n) =& \sum_{j=1}^{N_T} \sum_{l=0}^L c_{ij}(l) x_j^{PA}(n-l) \nonumber\\
  =& \sum_{j=1}^{N_T} \sum_{\substack{ p = 1 \\ p \text{ } \mathit{odd} }}^P
  \sum_{q = 0}^p \sum_{m = 0}^{M+L} \tilde{h}_{p,ij}^{(q,p-q)}(m)\times \label{eq:channel_output}\\
 & {x_j}{{(n-m)}^q}x_j^*{{(n-m)}^{p-q}} \nonumber
\end{align}
where $\tilde{h}_{p,ij}^{(q,p-q)}(m)={\mathop\sum}_{l=0}^{m} {c_{ij}}(l)h_{p,j}^{(q,p-q)}(m-l)$.

Eq. \eqref{eq:channel_output} does not yet include the effect of analog RF cancellation. As shown in Fig. 1, we assume \emph{active RF cancellation} where the PA output of each TX is tapped, and subtracted from each of the received signals after suitable gain, phase and delay adjustments. The RF cancellers can be either single-tap or multi-tap, but for generality we denote them with impulse responses $h_{ij}^{RF}(n)$, operating on the PA output signals $x_{j}^{PA}(n)$. Thus, the received SI signal of receiver $i$, after analog RF cancellation, becomes
\begin{align}
r_i(n) =& z_i(n) - \sum_{j=1}^{N_T} \sum_{l=0}^{L'} h_{ij}^{RF}(l) x_j^{PA}(n-l) \nonumber\\
  =& \sum_{j=1}^{N_T} \sum_{\substack{ p = 1 \\ p \text{ } \mathit{odd} }}^P
  \sum_{q = 0}^p \sum_{m = 0}^{M+\max(L,L')} \breve{h}_{p,ij}^{(q,p-q)}(m)\times \label{eq:received_si}\\
  & {x_j}{{(n-m)}^q}x_j^*{{(n-m)}^{p-q}}\text{.} \nonumber
\end{align}
with $\breve{h}_{p,ij}^{(q,p-q)}(m)={\mathop\sum}_{l=0}^{m} c_{ij}^{RF}(l)h_{p,ij}^{(q,p-q)}(m-l)$ and $c_{ij}^{RF}(l)=c_{ij}(l)-h_{ij}^{RF}(l)$. Here, $L'$ denotes the number of taps in the RF canceller. Notice that the structure of the model is still of the same form as in \eqref{eq:channel_output} but with modified impulse response coefficients and orders, taking into account the effect of the RF canceller.

Purely analog RF cancellation, as assumed in the above derivations, calls for ${N_T}\times {N_R}$ canceller circuits to be implemented in the device; one canceller from each transmitter to each receiver. The complexity may become prohibitive when the number of antennas is increased beyond 2. There is an alternative RF canceller structure, based on digital regeneration of the SI, and upconversion to RF via an additional transmitter chain \cite{Duarte12}, whose complexity scales with ${N_R}$ instead of ${N_T}\times {N_R}$. This type of a structure may prove to be more attractive with more antennas. We emphasize, however, that the proposed modeling is compatible with such RF cancellation as well, only the exact expressions of the model parameters, and their number (i.e., filter lengths), may be different. Hence, the structure of the SI signal model in \eqref{eq:received_si} is of general nature, independent of the actual RF cancellation implementation.

\subsection{Simplified Signal Models}
\label{sec:simplified}
With high transmit powers, and assuming typical RF component characteristics, the most powerful nonlinear SI terms in \eqref{eq:received_si} are (in approximate order): I/Q mismatch, PA 3rd order nonlinearity, PA 5th order nonlinearity, cascade of I/Q mismatch and PA 3rd order nonlinearity, PA 7th order nonlinearity, etc. The polynomial basis functions corresponding to these dominant self-interference terms are listed in Table \ref{table:si_terms}. In most practical cases all the other basis functions (except of course the linear basis function $x_j$) can be neglected. The relative strengths of the different SI terms naturally depend on the actual hardware of the device, as well as on the transmit power, so the relative order of some of the terms in Table \ref{table:si_terms} may eventually be different. The significant memory depths of the basis functions may also be different, even though in \eqref{eq:received_si} they are assumed to be the same for all the basis functions. In particular, the memory depths of the higher order (weaker) terms could be reduced without much effect on the overall modeling and cancellation accuracy. These practical insights on the RF components and their modeling allow simplifying the estimation and cancellation processing, and therefore alleviate the complexity increase associated with such joint modeling.

The receiver I/Q imbalances were not taken into account explicitly in the above modeling. In spite of this, it turns out that the obtained model can indeed model receiver I/Q imbalances as well. In practice, receiver I/Q imbalance would induce another widely-linear transformation on the received SI signal in \eqref{eq:received_si}, of the form $K_{1,i}r_i+K_{2,i}r^*_i$, where $K_{1,i}$ and $K_{2,i}$ are the receiver I/Q imbalance parameters, defined similarly as the transmitter counterparts below \eqref{eq:iq_model}. As the model in \eqref{eq:received_si} already contains all the possible complex-valued basis functions, none would be added by introducing receiver I/Q imbalance. Therefore, \eqref{eq:received_si} has the structural capability to model also receiver I/Q imbalances. In this case, however, the relative strengths of the terms given in Table \ref{table:si_terms} may change.

\begin{table}[!t]
\renewcommand{\arraystretch}{1.5}
\caption{Nonlinear self-interference terms in (approximate) order of strength.}
\label{table:si_terms}
\centering
\begin{tabular}{p{4.3cm}@{\hskip 10pt} | l}
%\hline
\textbf{Transmitter impairments} & \textbf{Basis functions}\\
\hline\hline
I/Q mismatch & $x_{j}^{*}$ \\[0.2em]
%\hline
PA 3rd order nonlinearity & $x_{j}^{2}x_{j}^{*}$\\[0.2em]
%\hline
Cascade of I/Q mismatch and PA 3rd order nonlinearity (stronger terms) & $x_{j}^{3}$, ${{x}_{j}}x_{j}^{*2}$ \\[0.2em]
%\hline
PA 5th order nonlinearity & $x_{j}^{3}x_{j}^{*2}$\\[0.2em]
%\hline
Cascade of I/Q mismatch and PA 3rd order nonlinearity (weak term) & $x_{j}^{*3}$ \\[0.2em]
%\hline
PA 7th order nonlinearity & $x_{j}^{4}x_{j}^{*3}$\\[0.2em]
%\hline
Cascade of I/Q mismatch and PA 5th order nonlinearity (stronger terms) & 	$x_{j}^{4}x_{j}^{*}$, $x_{j}^{2}x_{j}^{*3}$
%\hline
\end{tabular}
\end{table}
%\vspace{3mm}

\section{Self-Interference Parameter Estimation and Cancellation}
\label{sec:cancellation}

In general, digital self-interference cancellation in a FD MIMO device requires $N_T\times N_R$ responses (linear or nonlinear) to be estimated, one response from each transmitter to each receiver. These responses are then used to regenerate the self-interference and cancel it at the digital baseband of each receiver. Notice that, despite of certain similarity in the distortion basis functions, this processing is totally different compared to classical digital predistortion (DPD) of transmitters. In general, there are two basic approaches to digital cancellation of the nonlinear self-interference: (\textit{i}) build a complete linear-in-parameters model of the total SI signal (as in Section \ref{sec:modeling}), including the TX impairments, the linear MIMO channel, and RF cancellation, estimate the parameters of the developed model, and finally recreate and cancel the SI from the received signals; (\textit{ii}) have separate models for the linear MIMO channel and the transmitter impairments, estimate the parameters sequentially, and recreate and cancel the SI from the received signals. The latter approach has the benefit of smaller complexity, but a more elaborate estimation procedure is needed. In this article, we only concentrate on the linear-in-parameters approach, while the latter decoupled estimation and cancellation approach is considered an important future work item.

\begin{figure}[!t]
\centering
\includegraphics[width=\columnwidth]{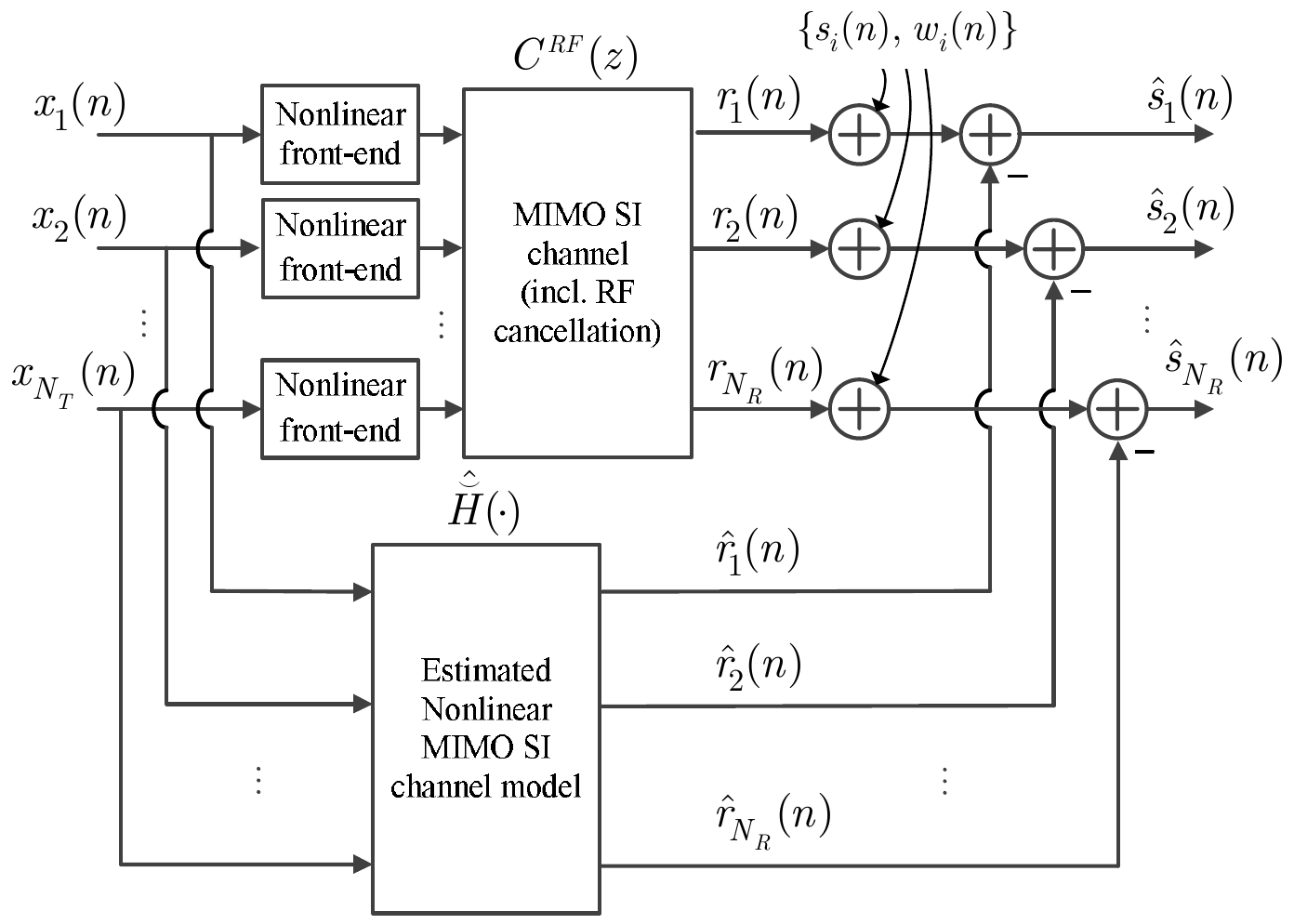}
\caption{Principle description of the proposed model for regeneration and cancellation of nonlinear self-interference.}
\label{fig:cancellation}
\end{figure}

\subsection{Proposed Self-Interference Canceller Structure}
The objective is now to estimate the parameters $\breve{h}_{p,ij}^{(q,p-q)}(m)$ based on the above SI signal model, and then to regenerate the SI signals and subtract them from the overall received signals at digital baseband. The self-interference replica generation and cancellation with the linear-in-parameters model follows \eqref{eq:received_si}, and is shown on a fundamental level in Fig. \ref{fig:cancellation}.

Formally, assuming that the desired signal of interest and additive noise at RX antenna $i$ (after propagating through the receiver) are denoted by $s_i(n)$ and $w_i(n)$, respectively, the overall received signal at digital baseband is written as
\begin{align}
	y_i(n) = r_i(n)+s_i(n)+w_i(n) \text{.}
\label{eq:rx_signal}
\end{align}
The output of the digital SI canceller at RX antenna $i$ is then
\begin{align}
	\hat{s}_i(n) = y_i(n) - \hat{r}_i(n) \text{,}
\label{eq:si_cancellation}
\end{align}
where $\hat{r}_i(n)$ denotes the nonlinear self-interference estimate, regenerated utilizing the model in \eqref{eq:received_si} and estimates of the model parameters $\breve{h}_{p,ij}^{(q,p-q)}(m)$.

\subsection{Parameter Learning}
Parameter learning based on least-squares (LS) estimation is outlined next. First, we write the vector/matrix representations of the relevant signals with \textit{N} observed samples as
\begin{align}
	\mathbf{y}_i &= \mathbf{r}_i + \mathbf{s}_i + \mathbf{w}_i \text{, with} \\
	\mathbf{y}_i &= \begin{bmatrix} y_{i}(n) & y_{i}(n+1) \cdots y_{i}(n+N-1) \end{bmatrix}^{\text{T}} \nonumber
\end{align}
and $\mathbf{r}_i$, $\mathbf{s}_i$, $\mathbf{w}_i$ are defined in the same manner as $\mathbf{y}_i$. Then we define the error vector as 
\begin{align}
	\mathbf{e}_i &= \mathbf{y}_i - \hat{\mathbf{r}}_i
\end{align}
where the nonlinear self-interference estimate is 
\begin{align}
	\hat{\mathbf{r}}_i &= \begin{bmatrix} \mathbf{\Psi}_1 & \mathbf{\Psi}_2 & \cdots & \mathbf{\Psi}_{N_T}  \end{bmatrix} \begin{bmatrix} \hat{\breve{\mathbf{h}}}_{i1}^T & \hat{\breve{\mathbf{h}}}_{i2}^T & \cdots & \hat{\breve{\mathbf{h}}}_{i N_T}^T \end{bmatrix}^T =\mathbf{\Psi}  \hat{\breve{\mathbf{h}}}_{i}  \text{.} 
\end{align}
Here, $\mathbf{\Psi}_j$ is a (horizontal) concatenation of the matrices
\begin{align}
\mathbf{\Psi}_{j,q,p}= \left[ \begin{smallmatrix} \psi_{j,q,p}(n) & \psi_{j,q,p}(n-1) & \!\cdots\! & \psi_{j,q,p}(n-\bar{M}+1) \\
\psi_{j,q,p}(n+1) & \psi_{j,q,p}(n) & \cdots & \psi_{j,q,p}(n-\bar{M}+2) \\
\vdots & \vdots & \ddots & \vdots \\
\psi_{j,q,p}(n+N-1) & \psi_{j,q,p}(n+N-2) & \cdots & \psi_{j,q,p}(n+N-\bar{M}) \\
\end{smallmatrix} \right]\label{eq:psi_matrix}
\end{align}
where $\psi_{j,q,p}(n)=x_j(n)^q x_j^*(n)^{p-q}$, with $j=1,2,\ldots,N_T$, $p=1,3,\ldots,\bar{P}$, and $q=1,2,\ldots,p$, and with $\bar{P}$ and $\bar{M}$ denoting the assumed polynomial order and memory depth of the overall nonlinear SI channel, respectively. The vector $\hat{\breve{\mathbf{h}}}_{ij}$ is a (vertical) concatenation of the vectors 
\begin{align}
	\hat{\breve{\mathbf{h}}}_{p,ij}^{(q,p-q)}& = \label{eq:h_vector}\\
	&[\hat{\breve{h}}_{p,ij}^{(q,p-q)}(0) ~\hat{\breve{h}}_{p,ij}^{(q,p-q)}(1) ~\cdots ~\hat{\breve{h}}_{p,ij}^{(q,p-q)}(\bar{M}-1)]^\text{T} \text{.} \nonumber
\end{align}
Depending on the choice of the relevant basis functions, all values of $q$ may not be used in \eqref{eq:psi_matrix}-\eqref{eq:h_vector} (refer to Table \ref{table:si_terms}).

The well-known least-squares solution to the parameter vector $\breve{\mathbf{h}}_{i}$ is then found as the solution which minimizes the power of the error vector $\mathbf{e}_i$, as %\cite{}
\begin{align}
	\hat{\breve{\mathbf{h}}}_{i} &= \operatorname*{arg\,min}_{\breve{\mathbf{h}}_{i}}  \left\| \mathbf{e}_i \right\|^2 = \operatorname*{arg\,min}_{\breve{\mathbf{h}}_{i}} \left\| \mathbf{y}_i - \mathbf{\Psi} \breve{\mathbf{h}}_{i} \right\|^2 \nonumber\\
	&= (\mathbf{\Psi}^\text{H}\mathbf{\Psi})^{-1} \mathbf{\Psi}^\text{H} \mathbf{y}_i \text{,}
\label{eq:ls_equations}
\end{align}
assuming full column rank in $\mathbf{\Psi}$.

Notice that for single-antenna full-duplex transceivers, in \cite{Korpi14} the authors propose a widely-linear SI canceller, i.e., one which employs the basis functions $x$ and $x^*$. Furthermore, in \cite{Anttila13} a nonlinear canceller which takes into account only the nonlinear basis functions induced by the PA ($x$, $x^2x^*$, $x^3x^{*2}$, etc.) is proposed. The joint MIMO nonlinear SI cancellation solution proposed in this paper can be seen to contain these single-antenna, single-impairment SI cancellers as special cases. In the next section, these SISO techniques (for the individual impairments) are adapted to MIMO, and compared against the proposed solution.

\section{Performance Simulations and Analysis}
\label{sec:simulations}
%SINR's with (1) conventional linear cancellation, (2) widely-linear cancellation, (3) PA nonlinearity cancellation, (4) proposed joint nonlinear cancellation ...

To determine the actual performance of the proposed algorithm in comparison to previous methods in \cite{Anttila13} and \cite{Korpi14}, we will evaluate them with full waveform simulations. The algorithms in \cite{Anttila13} and \cite{Korpi14} are for single-antenna FD devices, but they are adapted here to the MIMO scenario by considering the interference to each receive antenna from every transmit antenna. These extensions are easily obtained with the proposed formulation in \eqref{eq:received_si}, by fixing $q=(p+1)/2$ for the nonlinear canceller from \cite{Anttila13}, and by choosing $\bar{P}=1$ for the widely-linear canceller from \cite{Korpi14}. Notice that these special cases are novel as well, in the sense that no previous cancellation solutions taking into account either PA nonlinearity or I/Q imbalance in MIMO full-duplex devices exist in current literature.

\subsection{Simulation Parameters}
%The parameters of the simulated in-band full-duplex transceiver model are given in Table \ref{table:system_parameters}. The parameters of the relevant RF components of the transceiver are given in Table \ref{table:parameters}. Additional parameters for the simulator, mainly regarding the utilized OFDM waveform, are specified in Table \ref{table:simul_param}. 

The used waveform simulator follows the architecture shown in Fig.~\ref{fig:block_diagram} and it models each component explicitly using baseband equivalent models, which include the effects of all the considered circuit impairments without any approximations. These baseband equivalent models are constructed according to the parameters presented in Tables~\ref{table:system_parameters} and~\ref{table:parameters}, with additional parameters for the waveforms being shown in Table~\ref{table:simul_param}. Furthermore, the MIMO SI channel has a delay spread of 125 ns with Rayleigh fading impulse response taps, and a K-factor of 35.8 dB \cite{Duarte12}. For the PA, we consider two models. First, the system is simulated with a Hammerstein PA model, i.e., one which has a static nonlinearity followed by a linear filter \cite{Isaksson06}. With a Hammerstein PA model, the proposed modeling and cancellation can perfectly model the nonlinear SI arising from the cascade of I/Q imbalance and PA nonlinearity. The second PA model is Wiener, meaning that there is a linear filter preceding a static nonlinearity \cite{Schetzen81}. In this case there is a model mismatch between the PA model and the canceller structure. The mismatch is purposely included in order to demonstrate the robustness of the proposed solution under modeling inaccuracies.

The receiver nonlinearities are also modeled, with the parameters given in Table \ref{table:parameters}. Thus, in the simulations, all significant aspects of the full-duplex transceiver are modeled explicitly, including power amplifier nonlinearity, I/Q imbalance, analog SI cancellation, analog-to-digital converter (ADC) quantization noise, and receiver nonlinearities. In the simulations, 10000 samples are used to determine the canceller parameter estimates in each case, with the length of each impulse response $\breve{\mathbf{h}}_{p,ij}^{(q,p-q)}$ being 10. The signal of interest (SoI) is not present during estimation, even though the estimation can in principle be done with the SoI present as well. If included during the estimation, the SoI is seen as additional noise, thus increasing the estimation variance. The figure-of-merit used in the subsequent analysis is the signal-to-interference-plus-noise ratio (SINR) after digital SI cancellation. It is calculated based on the average value of 100 simulation runs at each transmit power level. The SoI is naturally present during SINR calculations, and it has a signal-to-noise ratio (SNR) of 15 dB. This level naturally represents the upper limit for the obtainable SINR. {\color{black}Also note that, since the target SNR is 10 dB, an SNR of 15 dB for the SoI indicates that it is only 5 dB above the receiver sensitivity level, resulting in a very challenging scenario in terms of SI cancellation.}

\begin{table}[!t]
\renewcommand{\arraystretch}{1.3}
\caption{System level and general parameters of the simulated 2x2 MIMO full-duplex transceiver.}
\label{table:system_parameters}
\centering
\begin{tabular}{|c||c|c|c||c|}
\cline{1-2} \cline{4-5}
\textbf{Parameter} & Value & & \textbf{Parameter} & Value\\
\cline{1-2} \cline{4-5}
SNR target & 10 dB & & RF cancellation & 30 dB\\
\cline{1-2} \cline{4-5}
Bandwidth & 12.5 MHz & & ADC bits & 12\\
\cline{1-2} \cline{4-5}
RX noise figure & 4.1 dB & & PAPR & 10 dB\\
\cline{1-2} \cline{4-5}
RX sensitivity & -88.9 dBm & & IRR (TX) & 25 dB\\
\cline{1-2} \cline{4-5}
RX input power & -83.9 dBm & & IRR (RX) & 50 dB\\
\cline{1-2} \cline{4-5}
Antenna separation & 40 dB & &  & \\
\cline{1-2} \cline{4-5}
\end{tabular}
\end{table}

\begin{table}[!t]
\renewcommand{\arraystretch}{1.3}
\caption{Parameters for the relevant components of the transmitter and receiver chains.}
\label{table:parameters}
\centering
\begin{tabular}{|c||c||c||c||c|}
\hline
\textbf{Component} & \textbf{Gain (dB)} & \textbf{IIP2 (dBm)} & \textbf{IIP3 (dBm)} & \textbf{NF (dB)}\\
\hline
PA (TX) & 27 & - & 13 & 5\\
\hline
LNA (RX) & 25 & - & 5 & 4.1\\
\hline
Mixer (RX)& 6 & 50 & 15 & 4\\
\hline
VGA (RX) & 0-69 & 50 & 20 & 4\\
\hline
\end{tabular}
\end{table} 

\begin{table}[!t]
\renewcommand{\arraystretch}{1.3}
\caption{Parameters for the OFDM waveform simulator.}
\label{table:simul_param}
\centering
\begin{tabular}{|c||c|}
\hline
\textbf{Parameter} & \textbf{Value}\\
\hline
Constellation & 16-QAM\\
\hline
Number of subcarriers & 64\\
\hline
Number of data subcarriers & 48\\
\hline
Guard interval & 16 samples\\
\hline
Sample length & 15.625 ns\\
\hline
Symbol length & 4 $\mu$s\\
\hline
Signal bandwidth & 12.5 MHz\\
\hline
Oversampling factor & 4\\
\hline
K-factor of the SI channel & 35.8 dB\\
\hline
\end{tabular}
\end{table}

\subsection{Results With Hammerstein PA Model}
\label{sec:hammerstein}
\begin{figure}[!t]
\centering
\includegraphics[width=\columnwidth]{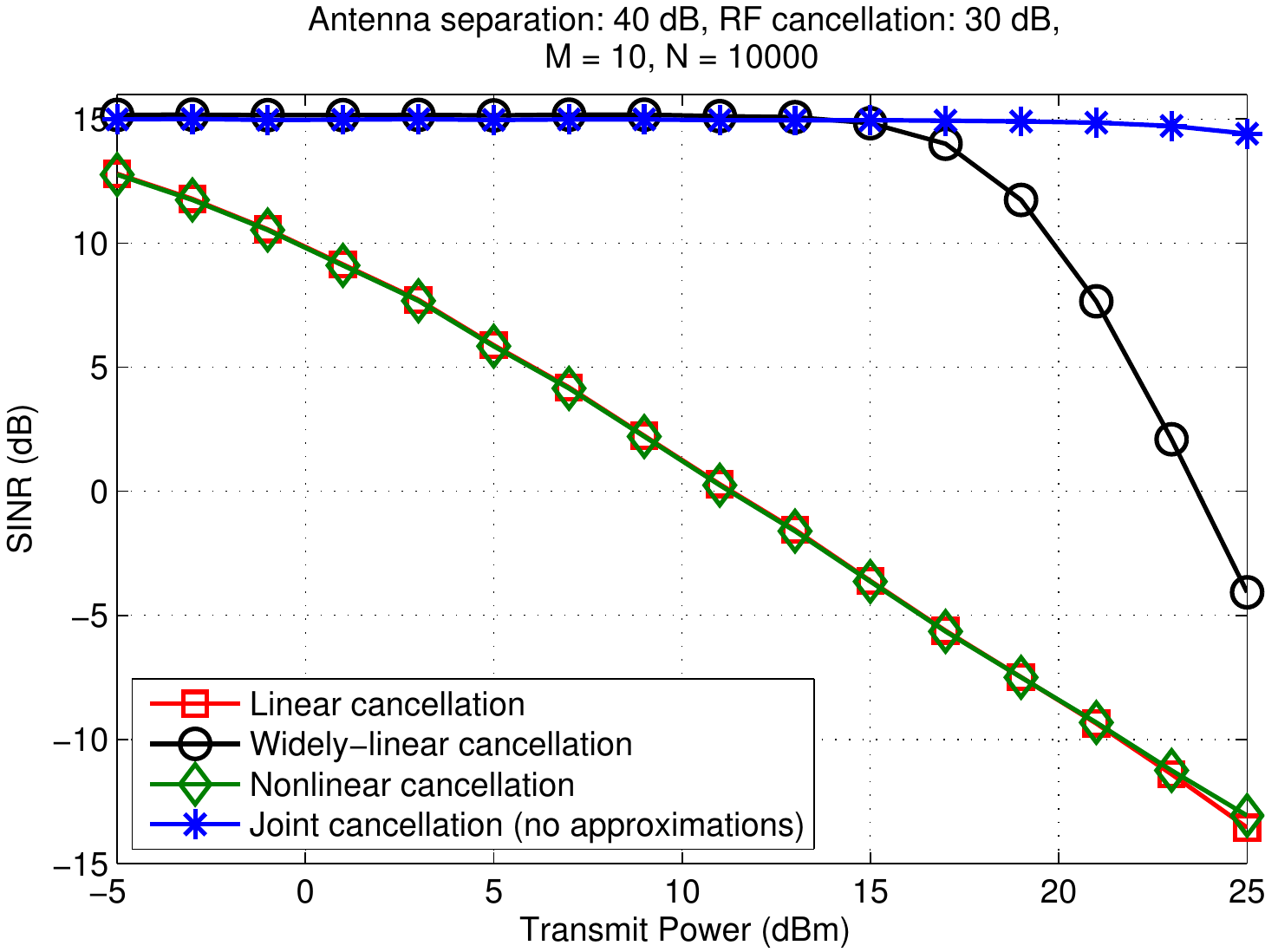}
\caption{The SINRs for different digital cancellation algorithms with respect to the overall transmit power. Hammerstein PA model.}
\label{fig:sinr_alg1}
\end{figure}

In these simulations, the memory of the Hammerstein PA is modeled with a 5-tap filter, and it is preceded by a 5th-order static nonlinearity. The resulting SINR curves for different algorithms, with respect to the total transmit power of the device, are shown in Fig.~\ref{fig:sinr_alg1}. In this example, the proposed joint nonlinear canceller contains all the basis functions for $\bar{P}=5$ and $\bar{M}=10$ defined by \eqref{eq:received_si}, i.e., no approximations are made regarding the corresponding signal model. It can be observed that performing only linear digital cancellation is clearly insufficient for achieving a high enough SINR in this scenario, except for the lowest transmit powers. The widely-linear digital cancellation, proposed in \cite{Korpi14}, is able to cancel the SI efficiently up to a transmit power of 15 dBm, after which its performance starts to decrease heavily. The reason for this is its inability to consider the PA-induced nonlinearities. Thus, when the SI signal is distorted nonlinearly more and more heavily, the widely-linear cancellation algorithm is not able to generate a sufficiently accurate cancellation signal. The nonlinear canceller from \cite{Anttila13}, on the other hand, can provide hardly any SINR improvement over the linear canceller in this scenario because it is not capable of modeling the I/Q imaging of the SI signal, which is dominating the overall interference profile at all power levels except the highest end.

However, as can be also observed from Fig.~\ref{fig:sinr_alg1}, the proposed joint canceller performs significantly better than either widely-linear cancellation or cancellation of only the PA-induced nonlinearities, since it takes both of these impairments into account \emph{by design}. With the joint cancellation algorithm, the SINR starts to decrease only with transmit powers above 21 dBm, and even then the drop is very mild. This drop has been determined to be caused by the receiver nonlinearities, which the proposed modeling cannot completely grasp{\color{black}, resulting in slightly higher levels of residual SI}. {\color{black}Also note that, with OFDM signals, the distribution of the residual SI is approximately multivariate Gaussian \cite{Shuangqing10}.}

\begin{figure}[!t]
\centering
\includegraphics[width=\columnwidth]{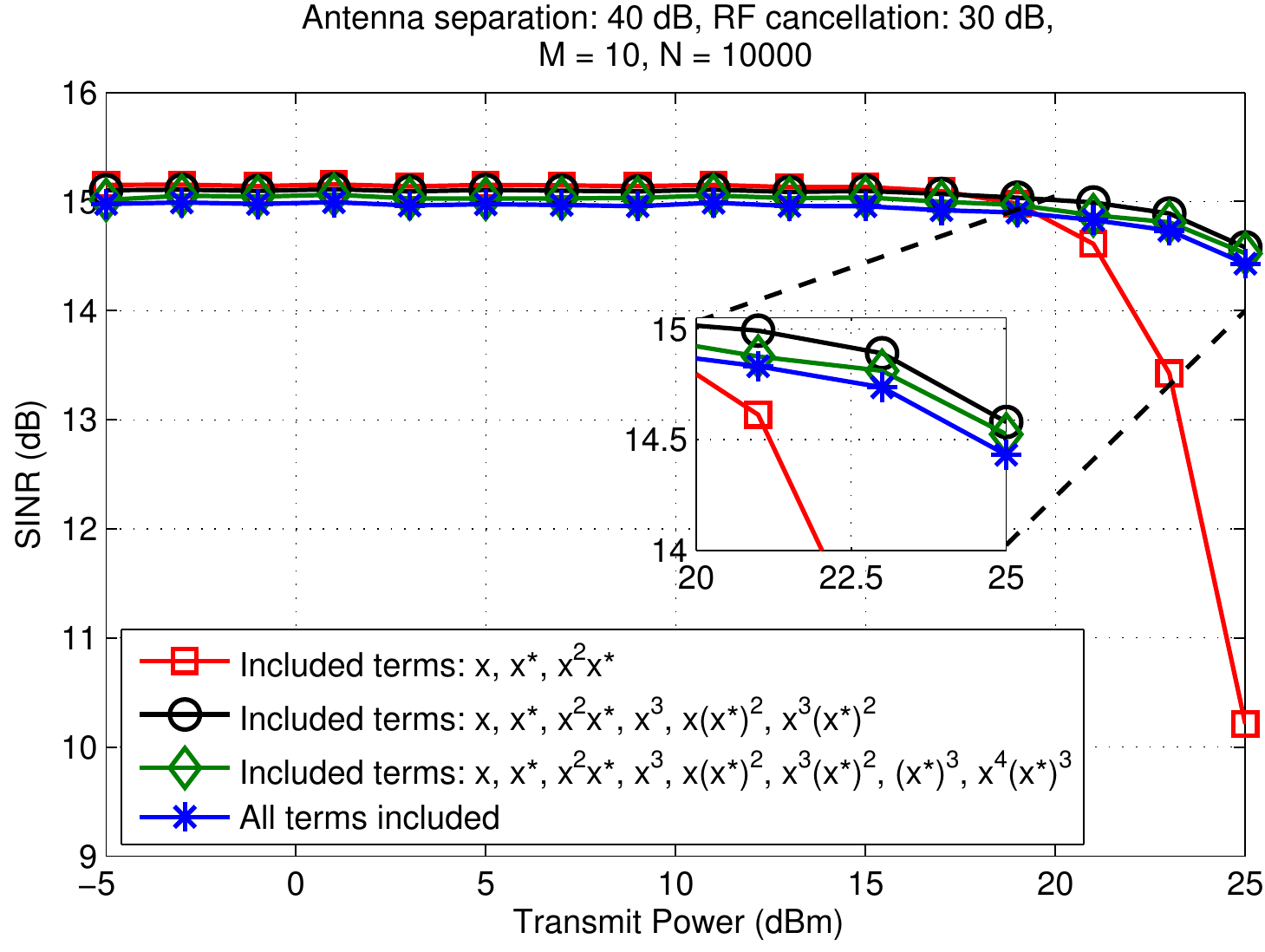}
\caption{The SINRs for different sets of included SI terms, listed in Table~\ref{table:si_terms}, with respect to the overall transmit power. Hammerstein PA model.}
\label{fig:sinr_appr1}
\end{figure}

In Table~\ref{table:si_terms}, the most dominant SI terms, stemming from the PA nonlinearities and I/Q imaging, are listed. To determine how much the different terms affect the actual estimation and cancellation performance, Fig.~\ref{fig:sinr_appr1} shows the SINRs for four different sets of basis functions, based on which the basis matrices are constructed. It is clear that including only the three most powerful terms is not sufficient to provide the optimal performance, and the SINR corresponding to this scenario falls approximately 4 dB below the highest achieved SINR at maximum transmit power. Furthermore, based on Fig.~\ref{fig:sinr_appr1}, it seems that the five most prominent terms account for nearly all of the nonlinear distortion with the considered component parameters. In fact, increasing the number of terms beyond five has a small negative effect on the SINR, as the estimation of a higher number of small parameters increases the variance of the final coefficients with a fixed sample size. {\color{black}This is a significant concern especially in higher order MIMO systems because the different terms must be generated for all the transmit signals, resulting in an even larger number of parameters that must be estimated.} However, it should be noted that the number of necessary terms depends heavily on the nonlinearity characteristics of the amplifier. For a more nonlinear PA, also the higher order terms might have a more significant effect on the accuracy of the cancellation signal. 

\subsection{Results With Wiener PA model}
\begin{figure}[!t]
\centering
\includegraphics[width=\columnwidth]{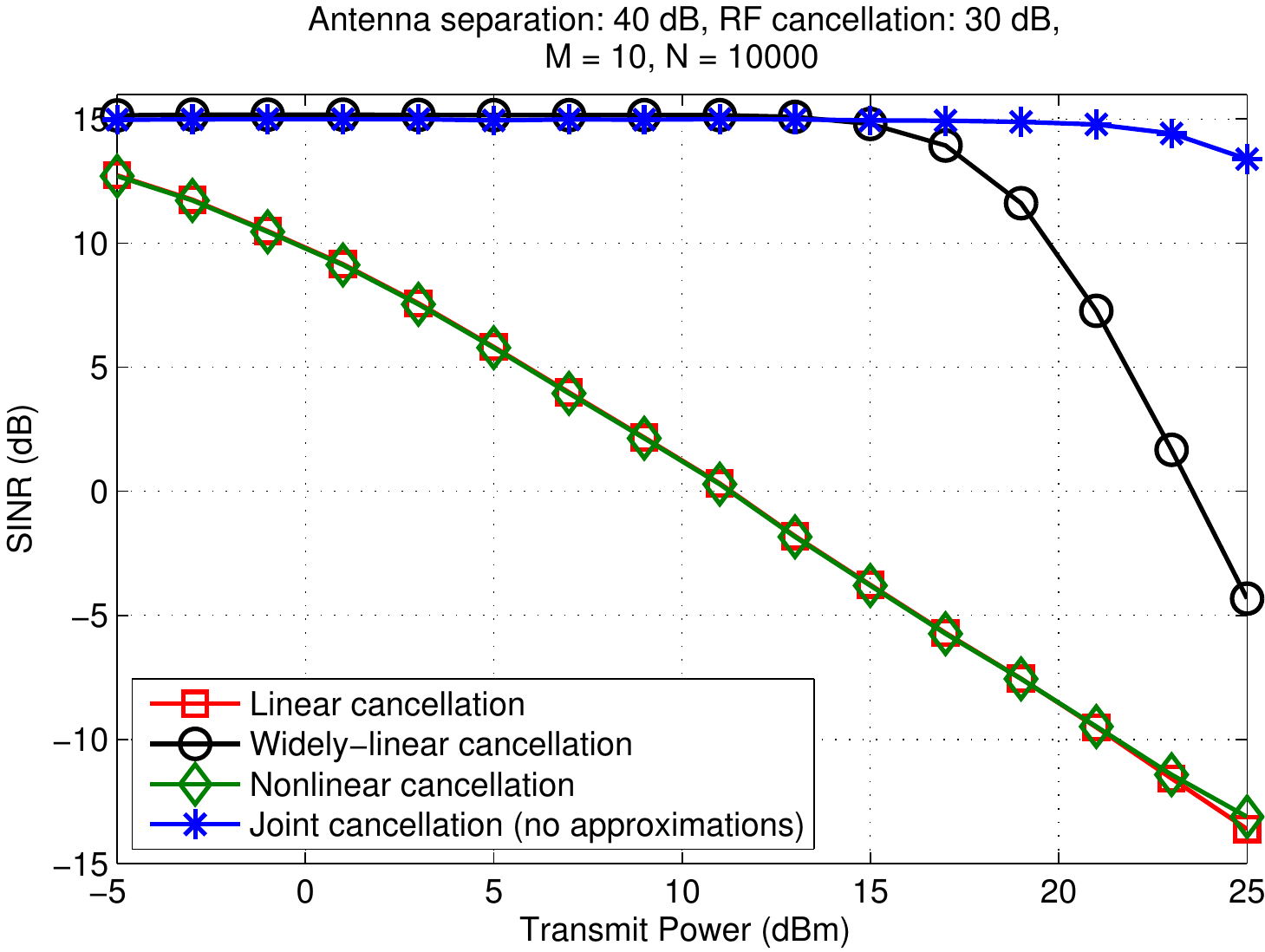}
\caption{The SINRs for different digital cancellation algorithms with respect to the overall transmit power. Wiener PA model.}
\label{fig:sinr_alg2}
\end{figure}

In Fig. \ref{fig:sinr_alg2} the simulations are repeated for the Wiener PA model, the simulation parameters being otherwise the same as above. The Wiener PA model filter has a length of 5 taps, the static nonlinearity order is 5, and the gain and IIP3 are as defined in Table \ref{table:parameters}. It can be observed from Fig. \ref{fig:sinr_alg2}, that the proposed canceller solution can provide a significant SINR improvement also under a Wiener PA model. The widely-linear canceller from \cite{Korpi14} and the nonlinear canceller from \cite{Anttila13} perform in almost exactly the same manner as with the Hammerstein PA model. The SINR achieved by the proposed solution starts to drop sooner and decreases more compared to the Hammerstein PA scenario, but the drop is still only about 1.5 dB at maximum transmit power. This drop is indeed mostly due to the model mismatch, i.e., having a Wiener PA model but assuming a parallel Hammerstein model in the canceller, even though, as concluded above, the receiver nonlinearities start to affect the obtainable SINR as well.

Overall, the results reported in this section clearly demonstrate that the proposed joint nonlinear SI canceller can efficiently suppress the self-interference under practical imperfect RF components, and it clearly outperforms the state-of-the-art reference techniques.

%\begin{figure}[!t]
%\centering
%\includegraphics[width=\columnwidth]{SINR_terms_v04.pdf}
%\caption{The SINRs for different sets of included SI terms, again with respect to the overall transmit power. Wiener PA model.}
%\label{fig:sinr_appr2}
%\end{figure}

\section{Conclusion}
\label{sec:conclusions}
Transceiver nonidealities, in particular power amplifier (PA) nonlinearity and I/Q modulator imbalance, have recently been shown to limit the performance and usable transmit power range of in-band full-duplex devices. In this article, detailed modeling of the self-interference in a full-duplex device with multiple transmit and receive antennas was carried out, including the effects of PA nonlinearity with memory and I/Q imbalance. Based on the obtained nonlinear self-interference model, a novel nonlinear spatio-temporal digital canceller was proposed. The proposed solution is the first of its kind in the sense that it takes into account both the dominant impairments, and is tailored for multiple antenna devices. The technique was shown by comprehensive simulations with two different PA models to clearly outperform the current state-of-the-art, being able to extend the usable transmit power range to over 20~dBm with practical, imperfect RF front-end models.

%\ifCLASSOPTIONcaptionsoff
%  \newpage
%\fi

%\newpage

\bibliographystyle{./IEEEtran}
\bibliography{./IEEEabrv,./IEEEref}

\end{document}